\documentclass[%
 reprint,
nofootinbib,
 amsmath,amssymb,
 aps,
]{revtex4-2}

\pdfoutput=1
\usepackage{graphicx}
\usepackage{tabularx}
\usepackage{rotating}
\usepackage{dcolumn}
\usepackage{bm}

\usepackage{amssymb}
\usepackage{pifont}
\usepackage{multirow}
\usepackage{mathrsfs}
\usepackage{mathtools}
\usepackage{makecell}
\usepackage{bbm}
\usepackage{amsmath}
\usepackage{bm}

\DeclarePairedDelimiter\ket{\lvert}{\rangle}
\DeclarePairedDelimiterX\braket[2]{\langle}{\rangle}{#1 \delimsize\vert #2}
\usepackage{mathtools}
\usepackage[toc,page]{appendix}

\usepackage{cellspace} %
\usepackage[utf8]{inputenc}

\usepackage[margin=1in]{geometry}
\usepackage{datetime}
\usepackage{xcolor}

\newdateformat{monthdayyeardate}{%
  \monthname[\THEMONTH]~\THEDAY, \THEYEAR}

\begin{document}

\title{Spontaneous Raman scattering out of a metastable atomic qubit}
\author{{I. D. Moore}, {A. Quinn}, {J. O'Reilly}, {J. Metzner}, {S. Brudney}, {G. J. Gregory}, {D. J. Wineland}, and {D. T. C. Allcock}}
\affiliation{Department of Physics, University of Oregon, Eugene, OR, USA}

\date{\monthdayyeardate\today}
\begin{abstract}
Metastable qubits in atomic systems can enable large-scale quantum computing by simplifying hardware requirements and adding efficient erasure conversion to the pre-existing toolbox of high-fidelity laser-based control. For trapped atomic ions, the fundamental error floor of this control is given by spontaneous Raman and Rayleigh scattering from short-lived excited states. We measure spontaneous Raman scattering rates out of a metastable $D_{5/2}$ qubit manifold of a single trapped $^{40}$Ca$^+$ ion illuminated by 976\,nm light that is -44\,THz detuned from the dipole-allowed transition to the $P_{3/2}$ manifold. This supports the calculation of error rates from both types of scattering during one- and two-qubit gates on this platform, thus demonstrating that infidelities $<10^{-4}$ are possible.
\end{abstract}

\maketitle


Stimulated Raman transitions are commonly driven to manipulate qubits encoded in trapped ions, neutral atoms~\cite{Yavuz2006,Jones2007}, defects in solid state systems~\cite{Yale2013,Goldman2020}, and more~\cite{Chou2020}. Raman-based quantum logic gates on trapped-ion qubits have been demonstrated with fidelities $\gtrsim99.9\%$~\cite{ballance2016,Gaebler2016}, sub-$\mu$s time scales~\cite{schafer_fast_2018}, absolute phase stability across space and time~\cite{Inlek2014}, straightforward individual qubit optical addressing~\cite{debnath_demonstration_2016}, and 3- and 4-body interactions~\cite{katz_demonstration_2023}. The fundamental error floor for these gates comes from spontaneous Raman and Rayleigh scattering through the short-lived $P_{1/2}$ and $P_{3/2}$ manifolds, which is already the limiting error source in the highest-fidelity experiments~\cite{ballance2016,Gaebler2016}. This limit can be reduced by increasing the Raman laser detuning while maintaining similar Rabi frequencies with a corresponding increase in laser power~\cite{MooreTheory}. Together with continued improvements to qubit and motional coherence, this will enable higher-fidelity operation of trapped-ion quantum computers and reduce the required overhead for fault-tolerant quantum computation.

Recent work~\cite{MooreTheory} expanded a model for spontaneous scattering rates of ground-state qubits in trapped ions~\cite{Ozeri2007} to also include qubits encoded in the metastable $D_{5/2}$ manifolds. Further, this theory extended beyond the small detuning regime by considering the frequency dependence of the density of states. The updated model has been experimentally verified for ground-state qubits in $^{133}$Ba$^{+}$~\cite{UCLAgBa} and predicts no fundamental error floor for ground-state qubits as the detuning between the Raman laser wavelength and the atomic transition grows, contrary to the model presented in Ref.~\cite{Ozeri2007}.

Encoding quantum information in metastable manifolds that are spectrally isolated from fluorescence operations enables mid-circuit recooling and measurement~\cite{omgBlueprint,Yang2022,shi2024long} and photonic interconnects for modularity~\cite{feng_realization_2024} without cumbersome dual-species operation~\cite{Moses2023,Drmota2023,OReilly2024}.
Furthermore, metastable qubits have a number of unique advantages, including heralded state preparation~\cite{sotirova2024high} and the efficient conversion of leakage errors into less harmful erasure errors~\cite{Wu2022,Kang2023,gatesLetter}. 

Recent demonstrations of entanglement between qubits encoded in the metastable $D_{5/2}$ manifolds of trapped barium~\cite{Wang2025,sotirova2024} and calcium~\cite{gatesLetter} ions have utilized laser-driven stimulated Raman transitions. Our work on high-fidelity logic gates in calcium~\cite{gatesLetter} took advantage of well-established laser technology at 976\,nm to achieve high optical power in a compact package and at a wavelength that is -44\,THz detuned from resonance with the dipole-allowed $D_{5/2}\leftrightarrow P_{3/2}$ transition at 854\,nm.

In this work, we measure the spontaneous scattering rate out of the metastable $D_{5/2}$ manifold of a $^{40}$Ca$^{+}$ ion induced by 976\,nm laser radiation. We confirm the theoretical prediction of Ref.~\cite{MooreTheory}, thus demonstrating that spontaneous Raman scattering errors can be suppressed to $<10^{-4}$ per two-qubit entangling gate at this wavelength. These conclusions are further supported by parallel work ~\cite{MITscattering} that measured scattering rates out of metastable barium ion qubits at multiple wavelengths.

\section{Experimental setup}
\label{sec:methods}
\subsection{Detection Scheme}
\label{sec:detection}

\begin{figure}
\centering
\includegraphics[width=\columnwidth]{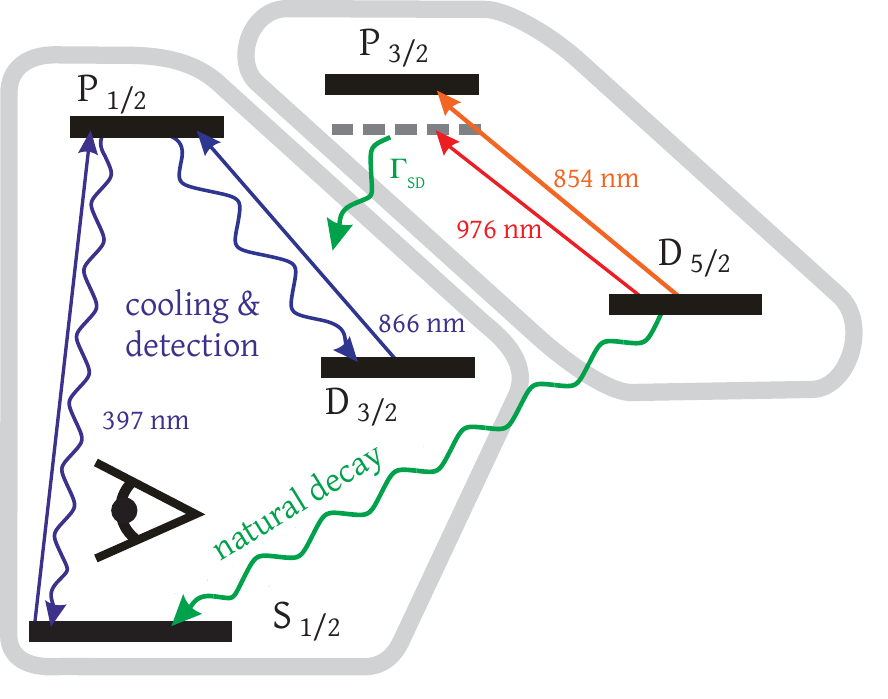}
\caption{Energy level diagram for $^{40}$Ca$^+$. The fluorescence cycle (blue) is spectrally isolated from the qubit manifold, enabling us to detect population outside the $D_{5/2}$ manifold using standard fluorescence collection techniques. We use this to measure decay (green) in the $D_{5/2}$ population due to its finite lifetime and scattering of the 976\,nm beam (red). Finally, we can depopulate $D_{5/2}$ by applying a resonant 854\,nm beam (orange).} 
\label{fig:leveldiagram}
\end{figure}

We trap a single $^{40}$Ca$^{+}$ ion in a four rod rf Paul trap~\cite{Metzner2024} and Doppler cool it by applying a 397\,nm beam with a projection onto all three principal axes of the trap and an 866\,nm beam blue-detuned from the $D_{3/2}\leftrightarrow P_{1/2}$ transition to repump out of the metastable $D_{3/2}$ manifold while avoiding dark resonances~\cite{Siemers1992}, see Figure~\ref{fig:leveldiagram}(a). We encode our qubit in the $\ket{\uparrow} \equiv \ket{D_{5/2},m_J=+5/2}$ and $\ket{\downarrow} \equiv \ket{D_{5/2},m_J=+3/2}$ metastable states and prepare the $\ket{\uparrow}$ state via polarization-selective optical pumping~\cite{gatesLetter}. A quantization field of 1.56\,G separates adjacent sublevels of the $D_{5/2}$ manifold by 2.63\,MHz via the linear Zeeman effect. All dipole-allowed transitions out of this manifold are detuned by at least several THz from the 397\,nm and 866\,nm beams; therefore an ion ``shelved" in $D_{5/2}$ will not scatter any light from the Doppler cooling beams~\cite{Keselman2011}. We can thus measure leakage out of $D_{5/2}$ by applying these beams and detecting 397\,nm fluorescence on a photo-multiplier tube, a technique we will refer to as a ``fluorescence check."

\begin{figure}
\centering
\includegraphics[width=0.8\columnwidth]{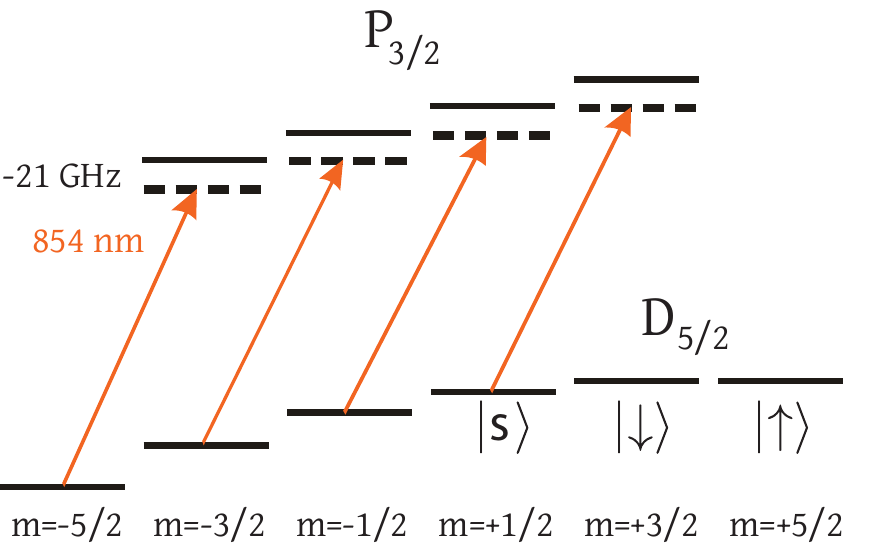}
\caption{Qubit definition and differential light shifting. The 2.63\,MHz separation between each $D_{5/2}$ sublevel is suppressed for clarity and other energies are not to scale. The $\sigma^+$-polarized 854\,nm beam -21.6\,GHz detuned from resonance with the $D_{5/2}\leftrightarrow P_{3/2}$ transition induces an AC Stark shift on each level proportional to its Clebsch-Gordan coefficient.} 
\label{fig:qubitdef}
\end{figure}

To prepare the $\ket{\downarrow}$ state after optically pumping to $\ket{\uparrow}$, we apply a near-resonant (-21.6 GHz) $\sigma^+$-polarized 854\,nm beam that breaks the $D_{5/2}$ manifold splitting degeneracy~\cite{sherman2013experimental}, see Fig.~\ref{fig:qubitdef}. Then, we use a resonant rf drive applied to one of the trap rods to generate an oscillating magnetic field and transfer population to $\ket{\downarrow}$. At the end of an experiment, we can distinguish between the qubit states by similarly driving population from $\ket{\downarrow}$ to $\ket{s} \equiv \ket{D_{5/2},m_J=+1/2}$ and applying a resonant $\pi$-polarized 854\,nm beam that pumps the $D_{5/2}$ sublevels with $|m|<5/2$ into the $S_{1/2}$ and $D_{3/2}$ manifolds with $>99\%$ fidelity. If the ion is dark during the subsequent fluorescence check, we apply another resonant 854\,nm beam with both $\sigma^+$ and $\sigma^-$ components to completely depopulate the $D_{5/2}$ manifold. If the ion is still dark after this, we assume that it either decrystallized\footnote{Significant heating events, most often collisions with background gas molecules, can disrupt the ion's coupling to laser beams and reduce detectable fluorescence to background levels.} or left the trap and we discard the measurement result.

\subsection{976\,nm Laser System}

\begin{figure}
\centering
\includegraphics[width=\columnwidth]{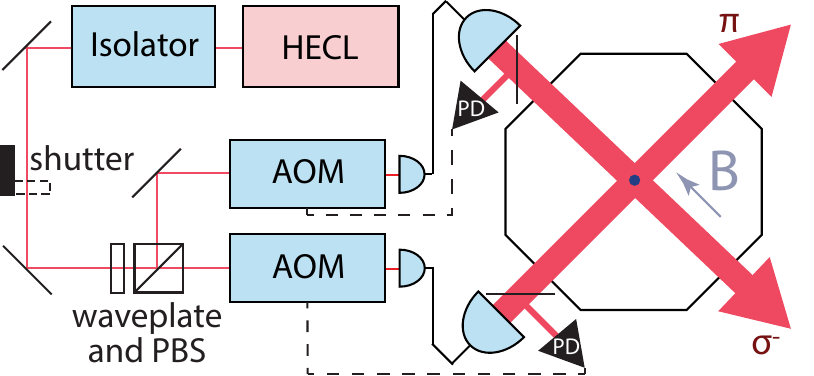}
\caption{976\,nm hybrid external cavity laser (HECL) system and beam geometry. The mechanical shutter is used to block or allow light to illuminate the ion. Only one beam path is used at a time in experiment.} 
\label{fig:laser_setup}
\end{figure}

Ref.~\cite{MooreTheory} predicts that two-qubit entangling gates can be performed on metastable calcium ions with $<10^{-4}$ error from spontaneous Raman scattering for any Raman transition wavelength greater than 963\,nm. As 976\,nm diode lasers are commonly used to pump erbium-doped fiber amplifiers for telecommunications, there are compact and high-power systems commercially available. We send 976\,nm light from a 700\,mW volume holographic grating-stabilized free-space HECL through an optical isolator and a mechanical shutter before it is split between two paths, see Fig.~\ref{fig:laser_setup}.

Each path corresponds to a different polarization at the ion, either $\sigma^-$ or $\pi$; we do not consider $\sigma^+$ light here as it does not drive any transitions from the qubit states to the $P_{3/2}$ manifold. We measured the scattering due to each polarization separately by adjusting the waveplate to direct all optical power into the corresponding path. The acousto-optic modulators (AOMs) in each path stabilize the beam powers based on feedback from photodiodes placed after beam samplers but before the vacuum chamber, see Figure~\ref{fig:laser_setup}. To maximize optical power at the ion, we use the 0th order of each AOM and achieve up to 180 and 220\,mW in the $\pi$ and $\sigma^-$ beams, respectively, at the trap and focus them to waists of $\sim30\,\mu$m and $\sim40\,\mu$m at the ion.

\subsection{Decay Rate Measurement}

We measure scattering out of $\ket{\uparrow}$ due to the $\sigma^-$-polarized beam and scattering out of $\ket{\downarrow}$ due to both the $\sigma^-$- and $\pi$-polarized beams. After state preparation, we illuminate the ion with a variable amount of 976\,nm light for a variable amount of time before performing the measurement sequence described in Section~\ref{sec:detection}, which distinguishes between electronic population in $\{S_{1/2},D_{3/2}\}$, $\{D_{5/2}\}-\ket{\uparrow}$, $\ket{\uparrow}$, or a missing ion. Each beam intensity is independently calibrated by measuring the AC Stark shift it induces on the $\ket{\uparrow}\leftrightarrow\ket{\downarrow}$ transition.

At a given intensity, we measure the $D_{5/2}$ population after several delay times up to 1\,s and fit this data to
\begin{equation}
    \textrm{Pop}_{D_{5/2}}(t)=e^{-t/\tau_{\textrm{meas}}}
\end{equation}
to extract the decay constant $\tau_{\textrm{meas}}$. For each combination of initial state and polarization, we also measure the populations over time with the shutter closed and find results consistent with the 1168(9)\,ms natural lifetime of the $D_{5/2}$ manifold~\cite{Kreuter2005}. The spontaneous Raman scattering rate out of $D_{5/2}$ is then
\begin{equation}
    \Gamma_{SD}=\frac{1}{\tau_{\textrm{meas}}}-\frac{1}{\tau_{\textrm{nat}}}
\end{equation}
where $\tau_{\textrm{nat}}$ is the measured natural lifetime.

\section{Results and Analysis}
\label{sec:results}
\begin{figure*}
\centering
\vspace{-7mm}
\includegraphics[width=\textwidth]{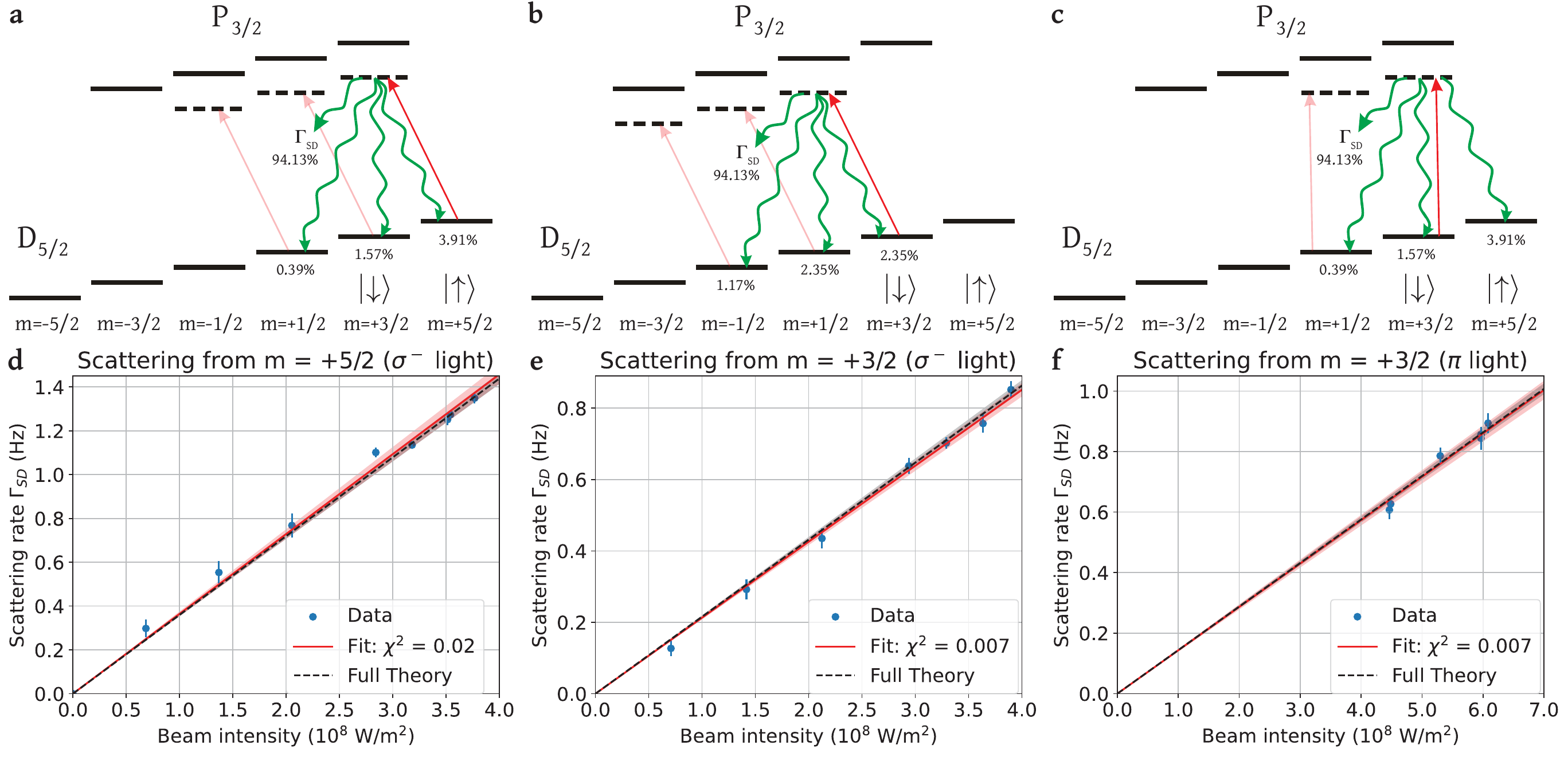}
\caption{Branching ratios and scattering rates for different initial states and beam polarizations. (a-c) Level diagrams for $\sigma^-$ scattering from $m=+5/2$ and $m=+3/2$ and $\pi$ scattering from $m=+3/2$, respectively. Energy separations are not to scale. The dark red arrows indicate the main laser driving and faded red arrows are pathways for second excitations after a back-scattering event. The green wavy arrows denote different destinations with decay to the $S_{1/2}$ and $D_{3/2}$ manifolds (not shown) represented by $\Gamma_{SD}$. (d-f) Scattering rates to $S_{1/2}$ and $D_{3/2}$ versus 976\,nm beam intensity with data (blue dots), a linear fit (red line) and theory prediction~\cite{MooreTheory} (gray dashed line). Shaded regions represent uncertainties in the fit and theory.} 
\label{fig:measurements}
\end{figure*}

For the case of $\sigma^-$ scattering from $\ket{\uparrow}$, we measured the scattering rates to $S_{1/2}$ and $D_{3/2}$ shown in Fig.~\ref{fig:measurements}(d). These could be inflated by events where the ion scatters back to $D_{5/2}$ and then to $S_{1/2}$ or $D_{3/2}$, but the branching ratio to $D_{5/2}$ is small ($5.88(3)\%$~\cite{Ramm2013}) and we expect less than one scattering event per trial on average for our delay times, so the effect should be small (amounting to an $\approx$0.6\% decrease in scattering rate). Indeed, we find agreement between the fitted and theoretical values~\cite{MooreTheory} to within error bars, given in Table~\ref{tab:results}, of scattering rate per unit of beam intensity. The fit uncertainty is statistical and the theory uncertainty is dominated by the $0.6\%$ uncertainty of the $P_{3/2}$ lifetime~\cite{Meir2020}.

\begin{table}[b]
\renewcommand{\arraystretch}{1.1}
\caption{\label{tab:results}
Fitted (see Fig.~\ref{fig:measurements}) and theoretically predicted~\cite{MooreTheory} spontaneous Raman scattering rates  to the $S_{1/2}$ and $D_{3/2}$ manifolds per unit 976\,nm intensity (Hz/(nW/m$^2$)) for different beam polarizations and initial states in $D_{5/2}$.}
\begin{ruledtabular}
\begin{tabular}{p{.1\textwidth}p{.1\textwidth}p{.1\textwidth}p{.1\textwidth}}
($m_J$, pol.) & (+5/2, $\sigma^-$) & (+3/2, $\sigma^-$) & (+3/2, $\pi$) \\
\colrule
Fit & 3.65(11) & 2.13(5) & 1.43(4)\\
Theory & 3.60(6) & 2.16(4) & 1.44(3) \\
\end{tabular}
\end{ruledtabular}
\end{table}

When we instead prepare the ion in $\ket{\downarrow}$ and apply $\sigma^-$-polarized light, we obtain the $SD$ scattering rates displayed in Fig.~\ref{fig:measurements}(e). We expect a larger double-scattering contribution of $1.4\%$ in this case, which is still within both our experimental and theoretical uncertainties. Finally, we again prepare the ion in $\ket{\downarrow}$ but apply $\pi$-polarized light, with results displayed in Fig.~\ref{fig:measurements}(f). In this configuration, scattering into $\ket{\uparrow}$ can have a larger effect on the measured $SD$ scattering rate as it will not scatter at all under the influence of the $\pi$ beam. To mitigate this effect, we simply disregard trials where we detect $\ket{\uparrow}$ population and again find agreement between theory and experiment. These results are summarized in Table~\ref{tab:results}.

\subsection{Gate Errors}
We now consider the scattering-induced errors in a representative single-qubit gate, the $\hat{\sigma_x}$ gate, driven by the 976\,nm lasers. The probability of a spontaneous Raman scattering error is~\cite{Ozeri2007}
\begin{equation}
P_{1q\textrm{Ram}}=\tau_{1q}\Gamma_{\textrm{Ram}}=\frac{\pi\Gamma_{\textrm{Ram}}}{2|\Omega_R|}
\end{equation}
where $\tau_{1q}$ is the gate time and $\Omega_R$ is the two-photon Rabi frequency. We scale the measured $\Gamma_{SD}$ to the more general $\Gamma_{\textrm{Ram}}$ using known branching ratios. Assuming equal intensities of the $\sigma^-$ and $\pi$ beams, 
we predict an error probability, including leakage to $S_{1/2}$, $D_{3/2}$, or $D_{5/2}$ outside of the qubit subspace and bit flip errors, of $1.25(2)\times10^{-6}$ per gate. Given the low branching ratio from $P_{3/2}$ to $D_{5/2}$, we can upper bound the Rayleigh-induced decoherence at $<10^{-7}$~\cite{moore2023}.

If we drive a $\sigma_z\otimes\sigma_z$ entangling gate on two qubits with the most efficient beam configuration, namely counter-propagating beams with equal intensities and pure $\sigma^-$-polarization, the spontaneous Raman scattering probability during the gate is~\cite{MooreTheory}
\begin{equation}
P_{2q\textrm{Ram}}=\tau_{2q}\Gamma_{\textrm{Ram}}=\frac{2}{\eta}\frac{\pi\Gamma_{\textrm{Ram}}}{2|\Omega_R|}
\end{equation}
where $\eta=\sqrt{\frac{\hbar}{2m\omega}}$ is the Lamb-Dicke parameter characterizing the laser beam's coupling to motion and the factor of two comes from the fact that there are now two ions with equal scattering probabilities. With a secular frequency $\omega/2\pi=2$\,MHz, we find $P_{2q\textrm{Ram}}=5\times10^{-5}$. In addition, Rayleigh scattering can cause two-qubit gate errors via the recoil momentum kick, which we can bound to $<10^{-8}$.

\section{Conclusion}\label{sec:Conclusion}
In summary, we measured spontaneous Raman scattering rates of metastable qubit states encoded in the $D_{5/2}$ manifold of a trapped $^{40}$Ca$^+$ ion under the influence of 976\,nm light that is far (-44\,THz) detuned from the nearest dipole transition and has been used to drive high-fidelity two-qubit logic gates~\cite{gatesLetter}. Using Clebsch-Gordan coefficients and known branching ratios~\cite{Song2019}, this allows us to calculate spontaneous Raman and Rayleigh scattering rates back to the $D_{5/2}$ manifold. Our results validate the theory of Ref.~\cite{MooreTheory}, which contradicts the previously-accepted model at these large detunings and predicts a fundamental error floor below $10^{-4}$ in our system.

Our work provides a strategy to reduce the largest infidelity in leading gate demonstrations~\cite{ballance2016,Gaebler2016} and supports theory that elucidates how to further lower the error floor. This is important because reducing error probabilities beyond the threshold for a given quantum error correction code reduces the overhead required to achieve a target logical error rate~\cite{knill_quantum_2005}.

\textbf*{Acknowledgments}
We acknowledge useful discussions with M. J. Boguslawski, Z. J. Wall, S. R. Vizvary, M. Bareian, E. R. Hudson, and W. C. Campbell. This research is supported in part by the NSF through the Q-SEnSE Quantum Leap Challenge Institute, Award \#2016244 and the US Army Research Office under award W911NF-20-1-0037. The data supporting the figures in this article are available upon reasonable request from D.T.C.A.

\bibliography{refs}

\clearpage

\newpage

\end{document}